\documentclass[reprint, superscriptaddress, secnumarabic, amssymb, nobibnotes, aps, prl]{revtex4-2}

\pdfoutput=1

\usepackage{graphicx}
\usepackage{epstopdf}
\usepackage[T1]{fontenc}
\usepackage[utf8]{inputenc}
\DeclareUnicodeCharacter{2212}{~}
\DeclareUnicodeCharacter{0327}{~}
\usepackage[utf8]{inputenc}
\usepackage{amsbsy}
\usepackage{gensymb}
\usepackage[T1]{fontenc}
\usepackage{amsmath}
\usepackage{amssymb}
\usepackage{bbm}
\usepackage{physics}
\usepackage{braket}
\usepackage{xcolor}
\allowdisplaybreaks
\usepackage{graphicx}
\usepackage[colorlinks=true]{hyperref}  
\hypersetup{
    bookmarks=true,         
    unicode=false,          
    pdftoolbar=true,        
    pdfmenubar=true,        
    pdffitwindow=false,     
    pdfstartview={FitH},    
    pdftitle={New breathing pyrochlore CuAlCr$_4$S$_8$},    
    pdfauthor={S. Sharma}, 
    pdfsubject={},   
    pdfcreator={},   
    pdfproducer={}, 
    pdfkeywords={} {} {}, 
    pdfnewwindow=true,      
    colorlinks=true,       
    linkcolor=blue, 
    citecolor=blue,        
    filecolor=magenta,      
    urlcolor=blue           
} 
\usepackage[normalem]{ulem}

\begin{document}
    
\preprint{APS/123-QED}

\title{Synthesis, physical and magnetic properties of CuAlCr$_4$S$_8$: a new Cr-based breathing pyrochlore }

\author{S.~Sharma}
\affiliation{Department of Physics and Astronomy, McMaster University, Hamilton, Ontario L8S 4M1, Canada}
\author{M. Pocrnic}
\affiliation{Department of Physics and Astronomy, McMaster University, Hamilton, Ontario L8S 4M1, Canada}
\author{B. N. Richtik}
\affiliation{Department of Chemistry, University of Manitoba, Winnipeg R3T 2N2, Canada}
\author{C. R. Wiebe}
\affiliation{Department of Chemistry, University of Manitoba, Winnipeg R3T 2N2, Canada}
\affiliation{Department of Chemistry, University of Winnipeg, Winnipeg R3B 2E9, Canada}
\author{J.~Beare}
\affiliation{Department of Physics and Astronomy, McMaster University, Hamilton, Ontario L8S 4M1, Canada}
\author{J. Gautreau}
\affiliation{Department of Physics and Astronomy, McMaster University, Hamilton, Ontario L8S 4M1, Canada}
\author{J. P. Clancy}
\affiliation{Department of Physics and Astronomy, McMaster University, Hamilton, Ontario L8S 4M1, Canada}
\affiliation{Brockhouse Institute for Materials Research, McMaster University, Hamilton, ON L8S 4M1, Canada}
\author{J. P. C. Ruff}
\affiliation{Cornell High Energy Synchrotron Source, Cornell University, Ithaca, NY, 14853, USA}
\author{M.~Pula}
\affiliation{Department of Physics and Astronomy, McMaster University, Hamilton, Ontario L8S 4M1, Canada}
\author{Q. Chen}
\affiliation{Department of Physics and Astronomy, McMaster University, Hamilton, Ontario L8S 4M1, Canada}
\affiliation{Brockhouse Institute for Materials Research, McMaster University, Hamilton, ON L8S 4M1, Canada}
\author{Y. Cai}
\affiliation{TRIUMF, Vancouver, British Columbia V6T 2A3, Canada}
\affiliation{Quantum Matter Institute, The University of British Columbia, Vancouver, BC V6T 1Z4, Canada}
\author{S. Yoon}
\affiliation{TRIUMF, Vancouver, British Columbia V6T 2A3, Canada}
\affiliation{Department of Physics, Sungkyunkwan University, Suwon 16419, Korea}
\author{G.~M.~Luke}
\email[]{luke@mcmaster.ca}
\affiliation{Department of Physics and Astronomy, McMaster University, Hamilton, Ontario L8S 4M1, Canada}
\affiliation{TRIUMF, Vancouver, British Columbia V6T 2A3, Canada}

\date{\today}

\begin{abstract}
\begin{flushleft}
\end{flushleft}
We present the synthesis and physical properties of a new breathing pyrochlore magnet CuAlCr$_4$S$_8$ with the help of synchrotron x-ray diffraction (XRD), magnetization under ambient and applied hydrostatic pressure, heat capacity, and muon spin relaxation/rotation ($\mu$SR) measurements.  CuAlCr$_4$S$_8$ exhibits positive thermal expansion with concave upward temperature dependence.  We observed a sharp antiferromagnetic ordering transition of a purely magnetic nature at 20 K, which shifts by as much as 3.2 K on application of 600 MPa pressure. The breathing factor (B$_f$ = $J'/J$) in breathing pyrochlores can be an important parameter to tune the magnetic ground states of pyrochlore lattice. The breathing factor can be modulated through breathing ratio, the ratio of sizes of the two tetrahedra, by using different elements at A and A' sites in the breathing pyrochlore structure.  We find that CuAlCr$_4$S$_8$ has a breathing ratio of 1.0663(8), which is comparable to other sulfur breathing pyrochlores.
\end{abstract}
\maketitle

\section{\label{sec:level1}INTRODUCTION \protect\\ }
Geometrically frustrated magnetic systems have been studied extensively over the last three decades \citep{Gardner2010}. The frustration between magnetic moments at corner sharing triangular or tetrahedral units can give rise to exotic ground states such as spin liquid \citep{Molavian2007}, ordered and disordered spin ice \citep{Mirebeau2005,Harris1997}, spin glass freezing \citep{Gingras1997} and order by disorder \citep{Savary2012} to name a few. The chromium spinels with formulae ACr$_2$X$_4$ are a family of such compounds where magnetic chromium atoms sit on corner sharing tetrahedra allowing various novel states, including zero energy excitation mode in spin liquid phase, heavy fermionic behavior, zero field and field induced transitions \citep{Lee2010}. 

\begin{figure}[]
\includegraphics[width=85.5mm]{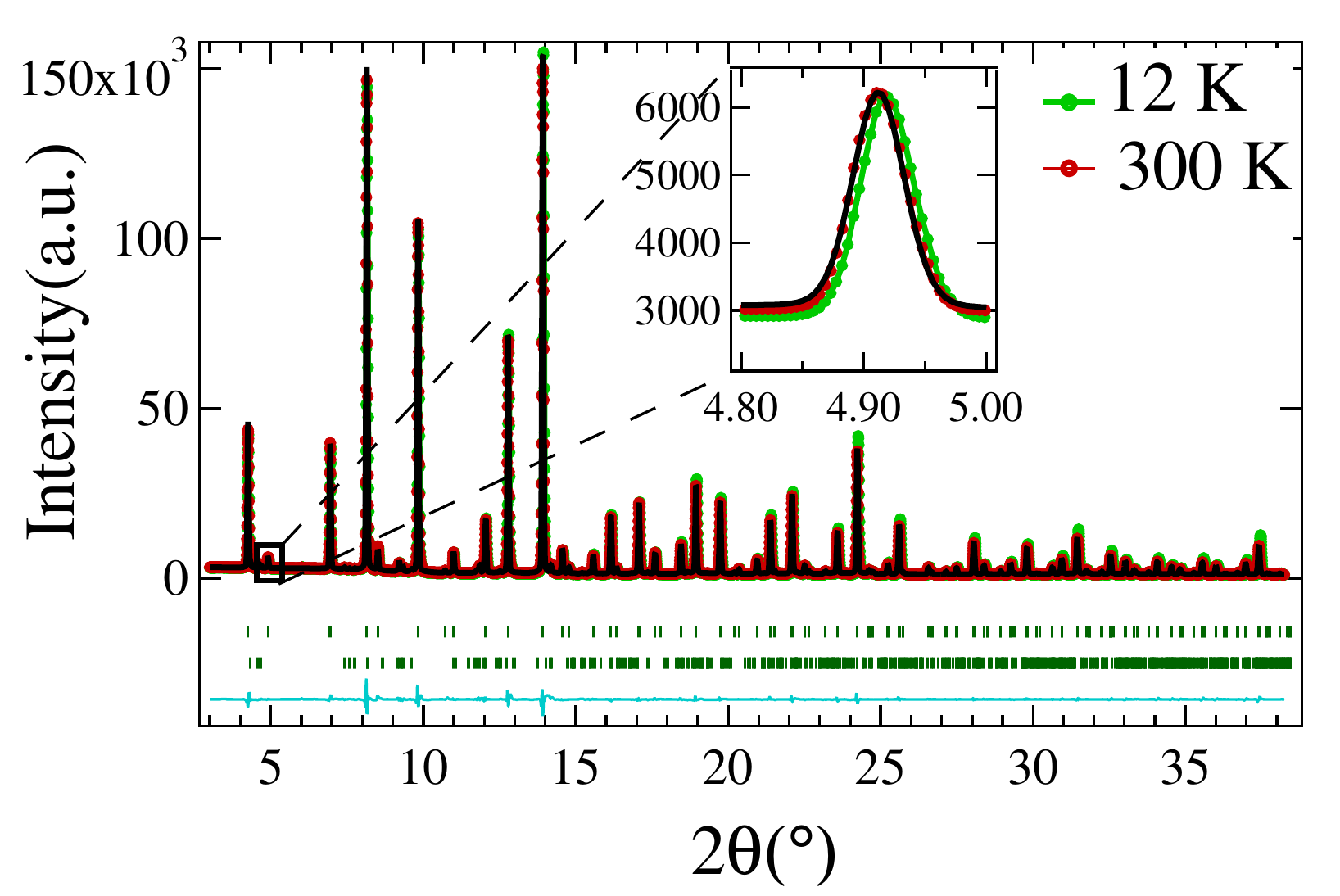}
\caption{\label{synchrotron} The figure shows the fit of the synchrotron powder x-ray diffraction data taken at 300 K and 12 K.  There is a small shift in the intensities and peak positions between the two temperatures, but no different peaks occur at the different temperatures. Inset shows the zoomed view of the [200] peak, characteristic of the breathing pyrochlore structure, presents evidence for a small shift in 2$\theta$ between the two scans. }
\end{figure}
The presence of two different types of non-magnetic elements at the A site leads to the formation of compounds with formulae AA'B$_4$X$_8$ and may lead to A-site ordering, which can allow the formation of  
larger and smaller tetrahedra. This occurs by displacement of the octahedral B-site cation and ordering of X anions \citep{Talanov2020}. Numerous mixed A-site pyrochlores have been synthesized in the past with some examples of the A-site ordering \citep[][]{Duda2008, Plumier1989, Plumier1977, Brasen1975, Wilkinson1976, Palmer1999, Lutz1989, Snyder2001, Snyder2000, Aminov2012}.   Since the type of interaction between nearest neighbor chromium atoms is distance sensitive, changing from strongly antiferromagnetic in oxide spinel to ferromagentic in selenium spinel \citep{Yaresko2008}, the magnetic Cr$^{+3}$ atoms at the edges of larger and smaller tetrahedra have different interaction (J and J'). This structure with alternating array of small and large tetrahera is called the Breathing Pyrochlore structure. Along with nearest neighbor interactions, further nearest neighbor interactions also play an important role in deciding the ground state \citep{Pokharel2020a, Ghosh2019}. This unique combination of magnetic frustration along with bond alternation in the breathing pyrochlores allow them to host exotic phases  including classical spin nematic transition in LiGa$_{0.95}$In$_{0.05}$Cr$_4$O$_8$ \citep{Wawrzynczak2017},  cluster frustration and negative thermal expansion in LiGaCr$_4$S$_8$ \citep{Pokharel2020a, Pokharel2018}, spin gap transition and magnetostructural transition in LiFeCr$_4$S$_8$ \citep{Saha2017}. These are a few examples of the diverse set of properties \citep{Reig-I-Plessis2021, Kalvius2008, Okamoto2015, Okamoto2013, Tanaka2018, Lee2021} exhibited by the breathing pyrochlores. 

Breathing degrees of freedom can be introduced in a regular spinel structure by substitution of A-site cation with nonequivalent cations. This reduces the symmetry of the lattice, converting centrosymmetric F$\bar{d}$3m (227) structure into non-centrosymmetric F$\bar{4}$3m (216), which belongs to the same group. This allows additional reflections like (200), (420), (600) etc.  in the diffraction pattern, which are forbidden otherwise.

We study the CuAlCr$_4$S$_8$ system that was first synthesized by H. L. Pinch et al. \citep{Pinch1970} in 1970, reporting the A-site ordering in the normal pyrochlore structure. We have  successfully synthesized CuAlCr$_4$S$_8$ in breathing pyrochlore structure. In this structure, non-magnetic Cu$^{+1}$ and Al$^{+3}$ ions  occupy the tetrahedral sites, whereas the Cr$^{+3}$ ions populate the octahedral sites, forming an array of corner sharing tetrahedra in the pyrochlore lattice. The Cr$^{+3}$ ion in Cr breathing pyrochlores adopts spin 3/2 state due to half-filled t$_{2g}$ orbital, which leads to the formation of an ideal spin 3/2 system with strong geometric frustration.

The Heisenberg spin Hamiltonian for a breathing pyrochlore can be written as \[H = J\sum_{ij}S_iS_j\] here J $\epsilon$ \{J, J$'$, J$_{nnn}$\}.  J and J$'$ are exchange coupling constants for small and large tertaherdra, respectively.  J$_{nnn}$ is interaction parameter between next nearest neighboring spins. The breathing factor (B$_f$ = J/J$'$) can be an important parameter to tune different ground states.

\section{EXPERIMENTAL DETAILS}
The polycrystalline sample of CuAlCr$_4$S$_8$ was synthesized by conventional solid-state synthesis method. We used high purity copper, aluminium, chromium and sulfur powders from Alfa Aesar in stoichiometric ratios and ground them into a pellet, which was sealed in a quartz tube under 10 mTorr of argon pressure. We heated the sealed tube at 1000  $^{\circ}$ C in a box furnace for 48 hours. This process was repeated at 1100 $^{\circ}$ C once more with intermediate grinding to obtain the pure sample. 

\begin{figure}[t]
\includegraphics[width=85.5mm]{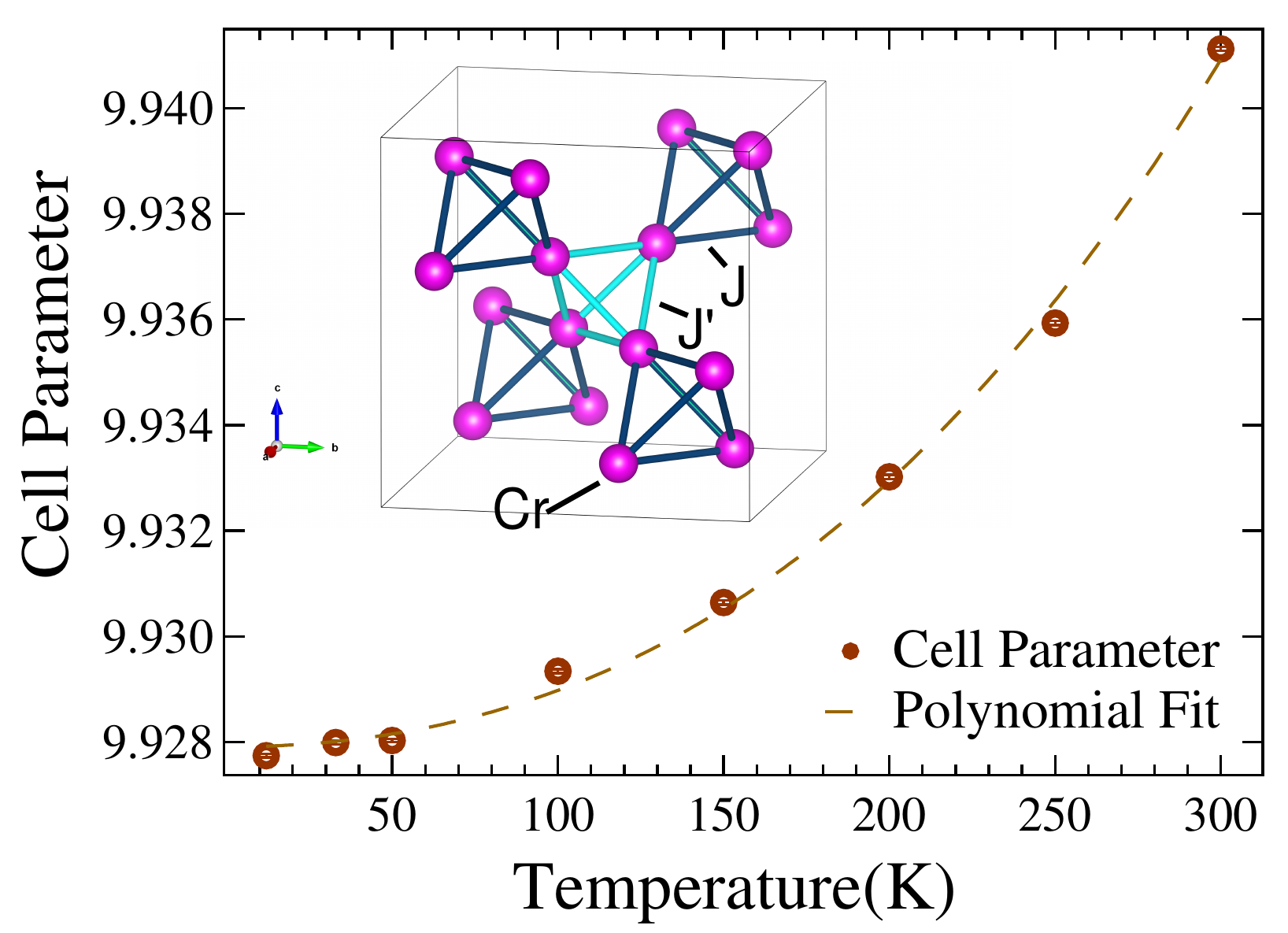}
\caption{\label{synchrotron} The figure shows the temperature dependence of the unit cell parameter obtained from the Rietveld refinement. The inset of the figure shows the breathing pyrochlore lattice formed by Cr atoms where the smaller and larger tetrahedra are distinguished through different colored bonds.}
\end{figure}

We performed preliminary room temperature powder X-ray diffraction measurements using a Panalytical X'pert diffraction instrument, which uses copper source with a wavelength of 1.54056 \AA. Rietveld refinement of the XRD pattern was performed using FullProf Suite. The sample contained 4.3 \% weakly ferromagnetic Cr$_3$S$_4$ \citep{Tazuke1979}, which does not interfere with our measurements except for magnetic susceptibility measurements in low fields. Magnetization measurements were performed using a Quantum Design SQUID with an RSO insert. We utilized a GC10/3 helium gas pressure cell from the Institute of High
Pressure Physics, Polish Academy of Sciences, inserted into
the MPMS to allow for magnetization measurements under
pressure up to 600 MPa, above 2 K. We performed specific heat measurements with a Quantum Design 9 T PPMS.
Synchrotron x-ray diffraction measurements were carried out on a powder sample of CuAlCr$_4$S$_8$ using the QM2 beamline at the Cornell High Energy Synchrotron Source (CHESS).  Measurements were performed using x-rays with an incident energy of 29.20 keV ($\lambda$ = 0.42460 \AA ~), and a Pilatus 6M area detector.  The sample was loaded into a borosilicate glass capillary (0.5 mm diameter), and cooled with a liquid helium cryostream.  Measurements were carried out from base temperature (T = 12 K) up to 300 K.  The zero and longitudinal field muon spin relaxation$\slash$rotation ($\mu$SR) measurements were performed at the M20 beamline at TRIUMF Vancouver using a He4 cryostate. The $\mu$SR data was fitted with the help of the musrfit software \citep{Suter2012}.
\begin{table}[]
\caption{\label{tab:table1}%
The crystallographic parameters of CuAlCr$_4$S$_8$ determined by Rietveld refinement of the synchrotron XRD data at 300 K. The space group is F$\bar{4}$3m. The lattice parameters are a = b = c = 9.94112(4) \AA\ and B is the thermal factor.
}
\begin{ruledtabular}
\begin{tabular}{c c c c c}
\textrm{ }&
\textrm{ }&
\textrm{x=y=z} & 
\textrm{B(\AA$^2$ )}& \textrm{Occupancy}\\
\colrule
 Cu1  &    4a &  0.75000(0)   &  0.451(58) &  0.994 \\
 Al1  &    4d &  0.00000(0) &  0.451(58) &  0.994 \\
 Al2  &    4a &  0.75000(0)  &  0.656(21) &  0.006 \\
 Cu2  &    4d &  0.00000(0)  &  0.656(21) &  0.006 \\
 Cr  &   16e &   0.37901(7)  &  0.431(11) &  1 \\
 S1   &   16e &  0.13212(5)  &  0.255(28) &  1 \\
 S2   &   16e &  0.61693(9)  &   0.431(27) &  1 \\

\end{tabular}
\end{ruledtabular}
\end{table}

\begin{table}[]
\caption{\label{tab:table2}%
The crystallographic parameters of CuAlCr$_4$S$_8$ determined by Rietveld refinement of the synchrotron XRD data at 12 K. The space group is F$\bar{4}$3m. The lattice parameters are a = b = c = 9.92774(3) \AA\ and B is the thermal factor.
}
\begin{ruledtabular}
\begin{tabular}{c c c c c}
\textrm{ }&
\textrm{ }&
\textrm{x=y=z} & 
\textrm{B(\AA$^2$ )}& \textrm{Occupancy}\\
\colrule
 Cu1  &    4a &  0.75000(0)   &  0.111(52) &  0.994 \\
 Al1  &    4d &  0.00000(0)  &  0.111(52) &  0.994 \\
 Al2  &    4a &  0.75000(0)  &  0.108(17) &  0.006 \\
 Cu2  &    4d &  0.00000(0)  &  0.108(17) &  0.006 \\
 Cr  &   16e &   0.37899(6)  &  0.130(10) &  1 \\
 S1   &   16e &  0.13217(5)  &  0.017(25) &  1 \\
 S2   &   16e &  0.61696(8)  &  0.129(24) &  1 \\

\end{tabular}
\end{ruledtabular}
\end{table}

\section{RESULTS AND DISCUSSION}
\paragraph*{Synchrotron X-ray diffraction}
The powder X-ray diffraction data taken on the polycrystalline samples of CuAlCr$_4$S$_8$ is shown in Figure \ref{synchrotron}. Rietveld refinement of the XRD data ensures good agreement with the breathing pyrochlore structure with space group F$\bar{4}$3m. This crystal structure carries magnetic Cr$^{+3}$ atoms at the corners of two unequal sized tetrahedra, unlike the normal pyrochlore (space group F$\bar{d}$3m) structure. The two Cr-Cr distances are 3.4020(14) and 3.6275(14) for this system. The ratio of the sizes of the two tetrahedra is called the breathing ratio, B$_r$, which is 1.0663(8) for CuAlCr$_4$S$_8$. This value is comparable to other sulfide breathing pyrochlores like CuInCr$_4$S$_8$ (1.06) \cite{Okamoto2018}, LiGaCr$_4$S$_8$ (1.077) \cite{Pokharel2018} and so on, however, it is much less then the rare earth compound Ba$_3$Yb$_2$Zn$_5$O$_{11}$ (1.90) \citep{Haku2016}, where the tetrahedra are decoupled.

\begin{figure}[b]
\includegraphics[width=85.5mm]{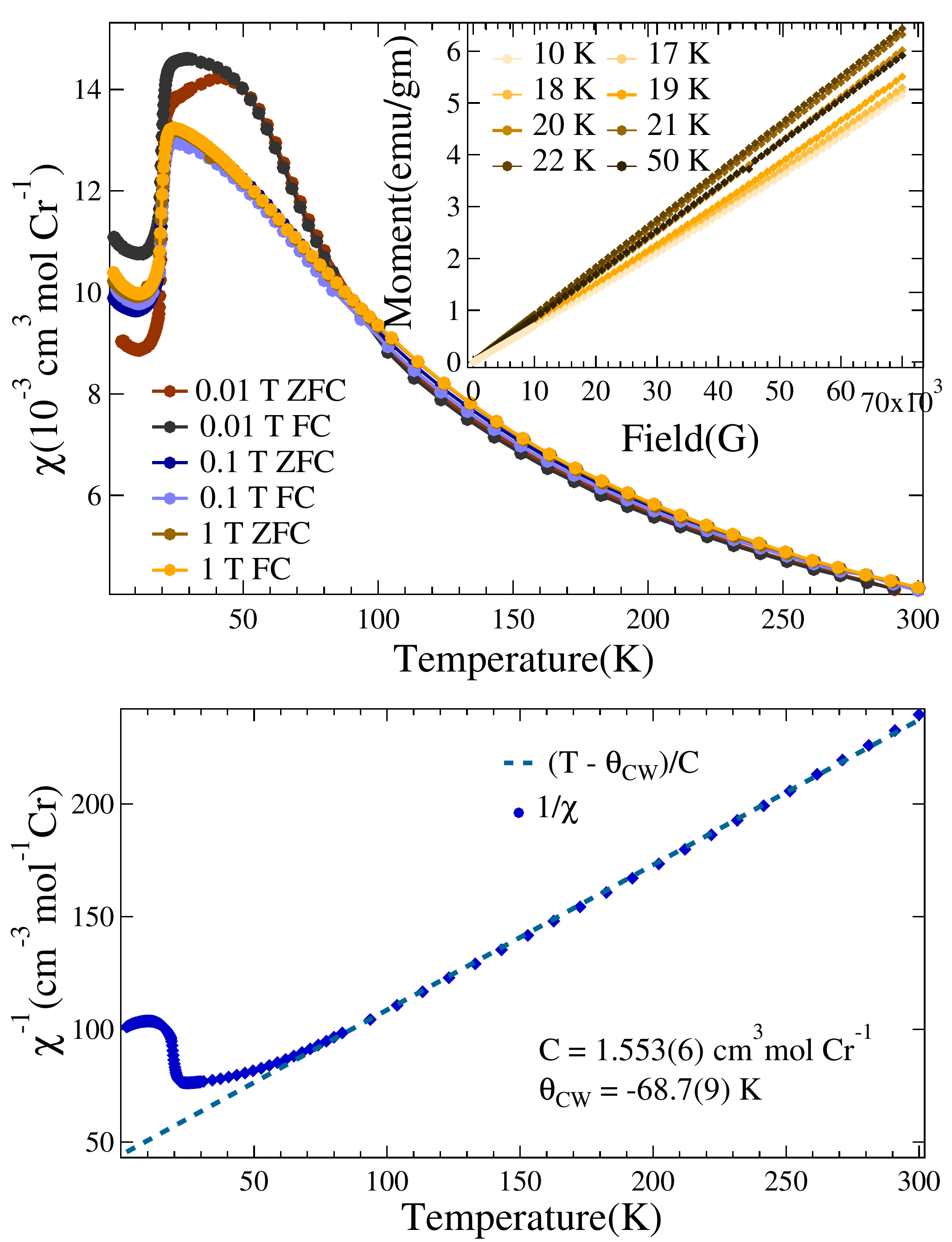}
\caption{\label{suscept}[Top]The dc magnetic susceptibility measurements on CuAlCr$_4$S$_8$. A sharp dip  at 20 K in the magnetic susceptibility, $\chi$, represents the antiferromagnetic transition. [Bottom] The inverse susceptibility measured in 0.1 T field and its Curie-Weiss fit in the temperature range 70 K-100 K. [Inset] The M-H at various temperatures around magnetic transition are plotted. }
\end{figure}

CuAlCr$_4$S$_8$ retains the room temperature crystal structure on cooling down to 12 K. The crystallographic parameters at 300 K and 12 K are listed in Tables \ref{tab:table1} and \ref{tab:table2}, respectively. The lattice parameters as a function of temperature, obtained from the Rietveld refinement, are shown in Figure: \ref{synchrotron}. It shows positive thermal expansion coefficient with a concave upward temperature dependence, unlike LiInCr$_4$S$_8$ and LiGaCr$_4$S$_8$, which show concave downward temperature dependence with positive thermal expansion and negative thermal expansion, respectively \citep{Okamoto2018}. This signifies the complex nature of the correlation between chemical composition and the structure of the breathing pyrochlore magnets.


\begin{figure}[b]
\includegraphics[width=85.5mm]{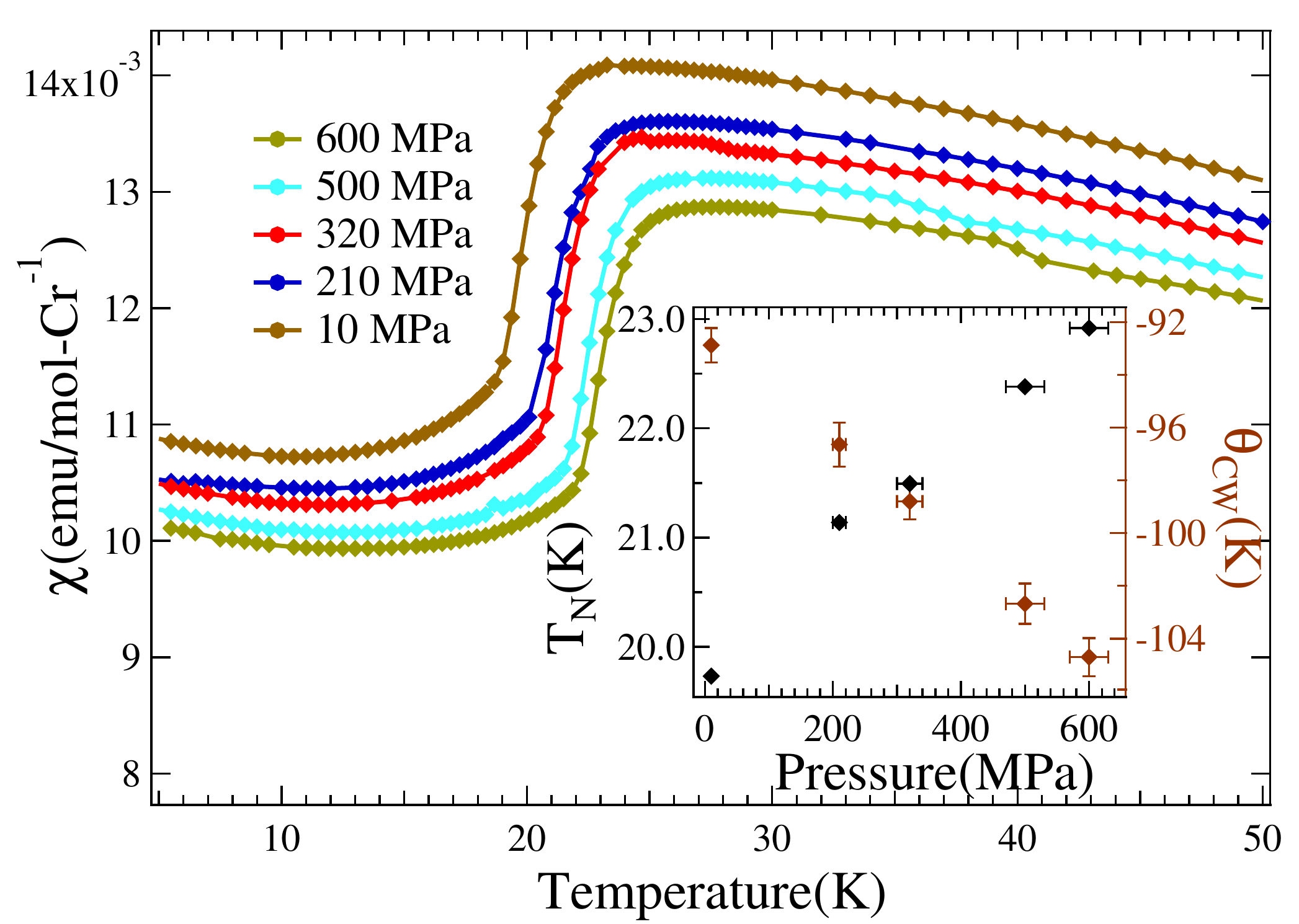}
\caption{\label{PressureSusceptibility} The graph shows the magnetic susceptibility plots measured with 0.1 T applied field under different pressures. [Inset] Shows the plot of midpoints of the magnetic transition and Curie-Weiss temperature against applied pressure}
\end{figure}


\paragraph*{Magnetization}
The dc magnetic susceptibility data in Figure \ref{suscept} shows a sharp anomaly at 20 K. We performed low temperature synchrotron XRD measurements, which found no evidence of the structure distortions associated the magnetic susceptibility anomaly. The negative Curie-Weiss temperature obtained from Curie-Weiss fit of the magnetic susceptibility establishes the dominant antiferromagnetic interaction in this system.   The antiferromagnetic ordering temperature, T$_N$, of CuAlCr$_4$S$_8$ is comparable to the other antiferromagnetic samples of this sulfide family like CuInCr$_4$S$_8$ (28 K), and  LiGaCr$_4$S$_8$ (10.3 K)   \citep{Okamoto2018}. 
The magnetic transition shifts to a higher temperature on application of hydro-static pressure as shown in the Figure \ref{PressureSusceptibility}. The shift in transition displays an almost linear dependence on pressure, with 600 MPa inducing a 3.2 K increase in the N\'eel temperature as shown in Figure \ref{PressureSusceptibility}[Inset].   This may be due to changing volume of the crystalline lattice induced by hydro-static pressure, which generates stronger interactions among the magnetic spins, allowing for higher magnetic ordering transition temperature. Magnetic susceptibility data were fitted with the Curie-Weiss equation in the range 70 K to 300 K as shown in Figure \ref{suscept} (b).  
\begin{equation}
\label{CWeq}
\chi(T) = \frac{C}{T-\theta_{CW}}  
\end{equation}
where $C = N_Ag^2\mu_B^{2}S(S+1)/3K_B$. The value of Curie-Weiss constant (C = 1.553(6)) obtained from the fits of the ambient pressure data corresponds to an effective magnetic moment ($\mu_{eff}$) of 3.525(7)*$\mu_B$, which is lower than the other sulfide breathing pyrochlores. This value of $\mu_{eff}$ yields a g = 1.83 for S = 3/2. The Curie-Weiss temperature ($\theta_{CW}$) for CuAlCr$_4$S$_8$ is -68.7(9) K. The smaller value of the N\'eel temperature in comparison to Weiss temperature signifies the suppressed antiferromagnetic (AFM) long-range interactions due to frustration. The frustration factor (T$_f$ = $\theta_{CW}/T_N$) for CuAlCr$_4$S$_8$ is 3.43(7). Curie-Weiss fit of the 600 MPa data exhibits the same Curie-Weiss constant (within the error limit) as ambient pressure data implying the unchanged effective moment. However, Curie-Weiss temperature  shifts to a lower temperature on application of pressure, the temperature dependence is plotted in \ref{PressureSusceptibility} [Inset]. Moreover, the frustration factor, the ratio of $\theta_{CW}$ and $T_N$, decreases by 2.9 \% at 600 MPa pressure.   
\begin{figure}[b]
\includegraphics[width=85.5mm]{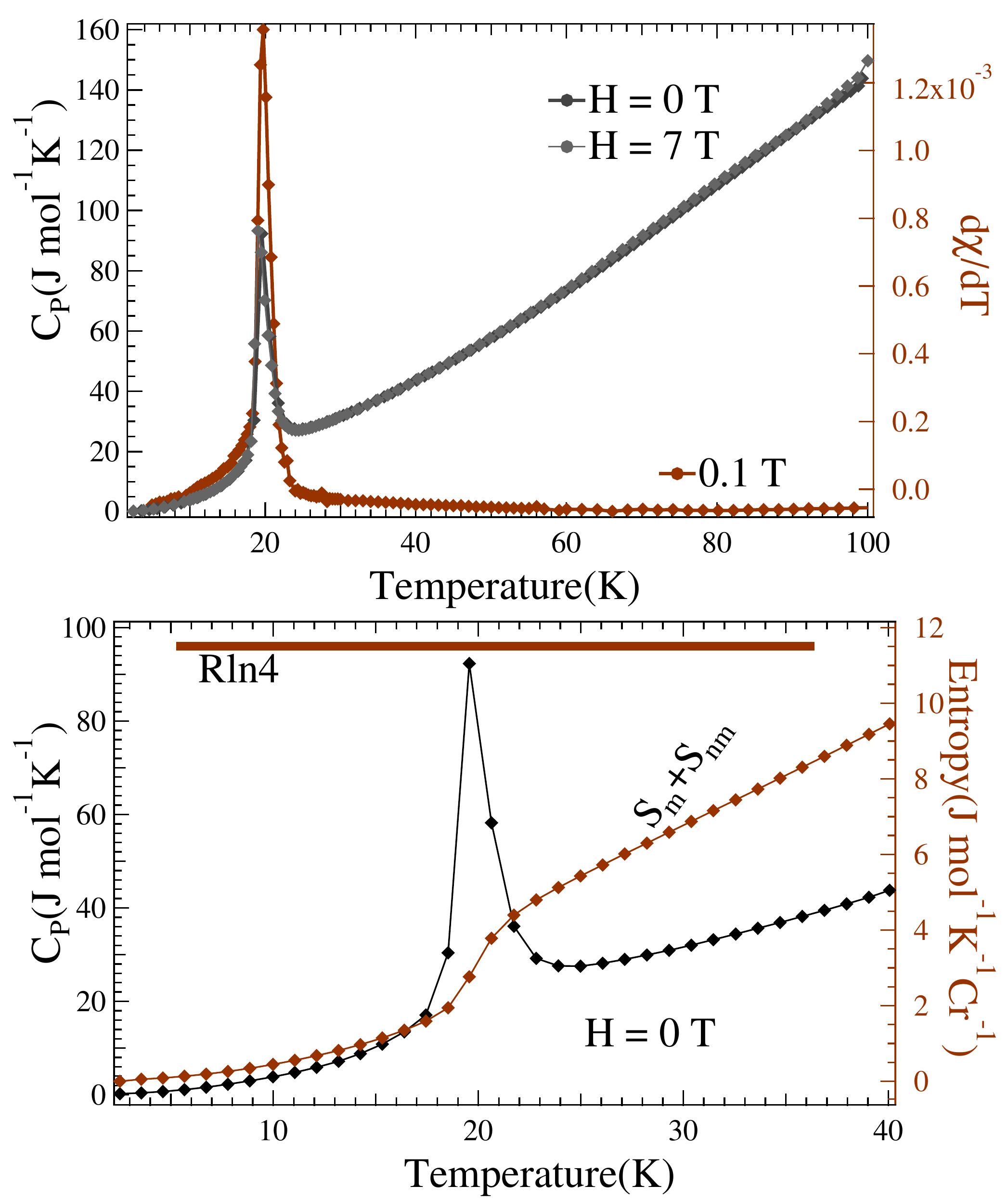}
\caption{\label{HeatCap} [Top] The figure shows that the sample heat capacity and the slope of the magnetic susceptibility peak at the same temperature. [Bottom] The expanded view of zero field heat capacity to elucidate the near T$_N$ ranges. The total entropy (magnetic + non-magnetic) and the maximum entropy of a spin 3/2 system are shown here for comparison.}
\end{figure}
\paragraph*{Heat Capacity} 
Heat capacity measurements performed on a polycrystalline sample of CuAlCr$_4$S$_8$ are shown in Figure: \ref{HeatCap}. There is a sharp peak in heat capacity data at 20 K,  which is largely independent of the applied field. The bottom panel of the figure depicts the total entropy as a function of temperature. Due to the absence of the non-magnetic analog of this sample, we could not subtract the lattice contribution from the entropy. The system is not able to recover the full entropy of a spin 3/2 magnetic system until 40 K, despite not correcting for the lattice contribution. The temperature derivative of magnetic susceptibility manifests the same behavior as specific heat near the magnetic transition Figure: \ref{HeatCap} [Top], which in principle looks antiferromagnetic; however, a detailed neutron diffraction experiment is desired to definitely determine the nature of this transition. 
\begin{figure}[b]
\includegraphics[width=85.5mm]{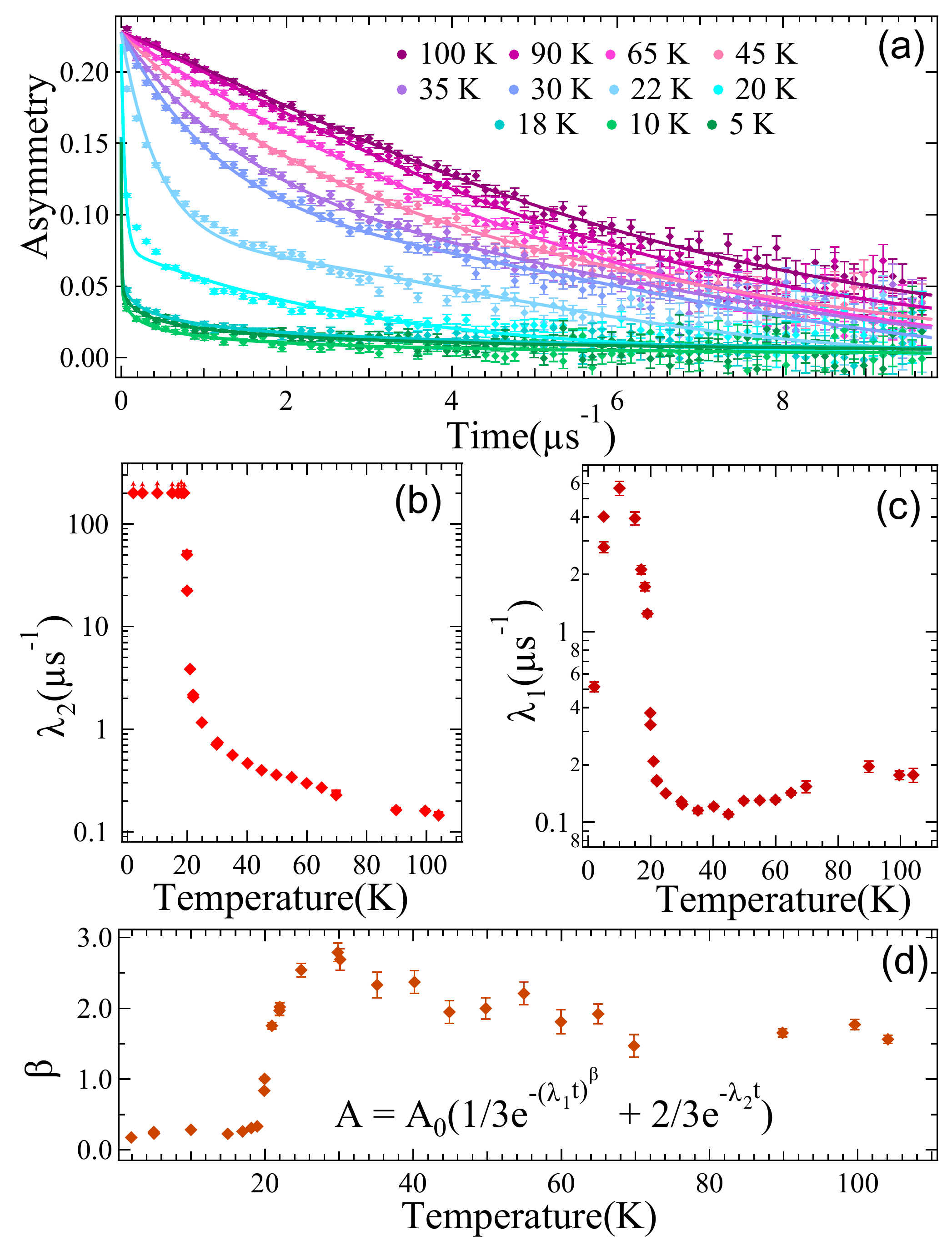}
\caption{\label{ZF_musr}  The zero field (ZF) $\mu$SR data taken on polycrystalline sample of CuAlCr$_4$S$_8$. (a) The asymmetry spectra taken at various temperatures above and below the T$_N$. The relaxation rates (b) $\lambda_1$ and (c) $\lambda_2$, and (d) stretched exponential factor, $\beta$  as obtained from the fits of the asymmetry are plotted against temperature. All three parameter shows an abrupt change of relaxation rates as we cool down the sample below its ordering temperature. The relaxation rate, $\lambda_2$, is too fast below 20 K and contained to be less than 200 $\mu s^{-1}$ and therefore poorly determined.}
\end{figure}

\paragraph*{Zero-field (ZF)/ longitudinal field (LF) muon spin relaxation/rotation ($\mu$SR)} 
Figure \ref{ZF_musr} (a) shows zero field $\mu$SR spectra taken at temperatures between 2 K to 100 K. The sample relaxation rate starts increasing as we cool down from 100 K and major part of the signal relaxes out on going through the magnetic transition. We do not see any spontaneous oscillations in the signal in the ordered state, such as was seen with LiGaCr$_4$O$_8$ \citep{Lee2016} and ZnCr$_2$O$_4$ \citep{Marshall2002}. The fast relaxing part of the signal is fitted with an exponentially relaxing function, whereas the 1/3rd tail is fit with a stretched exponential function as shown in equation \ref{asym} 
\begin{equation}
\label{asym}
A = \frac{A_0}{3}{e^{-(\lambda_1t)}}^{\beta}+\frac{2A_0}{3}e^{-\lambda_2t} 
\end{equation}
 where A is total asymmetry and $A_0$ is  initial asymmetry. $\lambda_1$ and $\lambda_2$ are fast and slow relaxation rates, respectively. $\beta$ is the exponent of the stretched exponential factor fitting to the 1/3rd tail of the signal. In a polycrystalline sample, the spins are pointing in random directions, 1/3 component of the signal, also called 1/3rd tail, describes the asymmetry due to the spins parallel to initial muon spin while the remaining 2/3rd part describes the spins perpendicular to initial muon spin direction. The monotonically increasing relaxation rates upon cooling signifies the slowing down of thermal fluctuations. Upon cooling below 20 K, the asymmetry signal starts to decay rapidly. Thus, the relaxation rates rise sharply, indicating the quasi-static magnetic ordering at 20 K. The absence of the oscillating component below the transition  means that the field distribution is very broad. The fact that we have 1/3rd and 2/3rd components means that the entire sample undergoes spin freezing. The internal field below T$_N$ = 20 K is very large and as such we are unable to resolve the time dependence of the muon polarization, either to detect oscillations as would be expected in an ordered state, or a non-oscillating but rapidly relaxing signal as would be expected in a glassy state.

\begin{figure}[t]
\includegraphics[width=85.5mm]{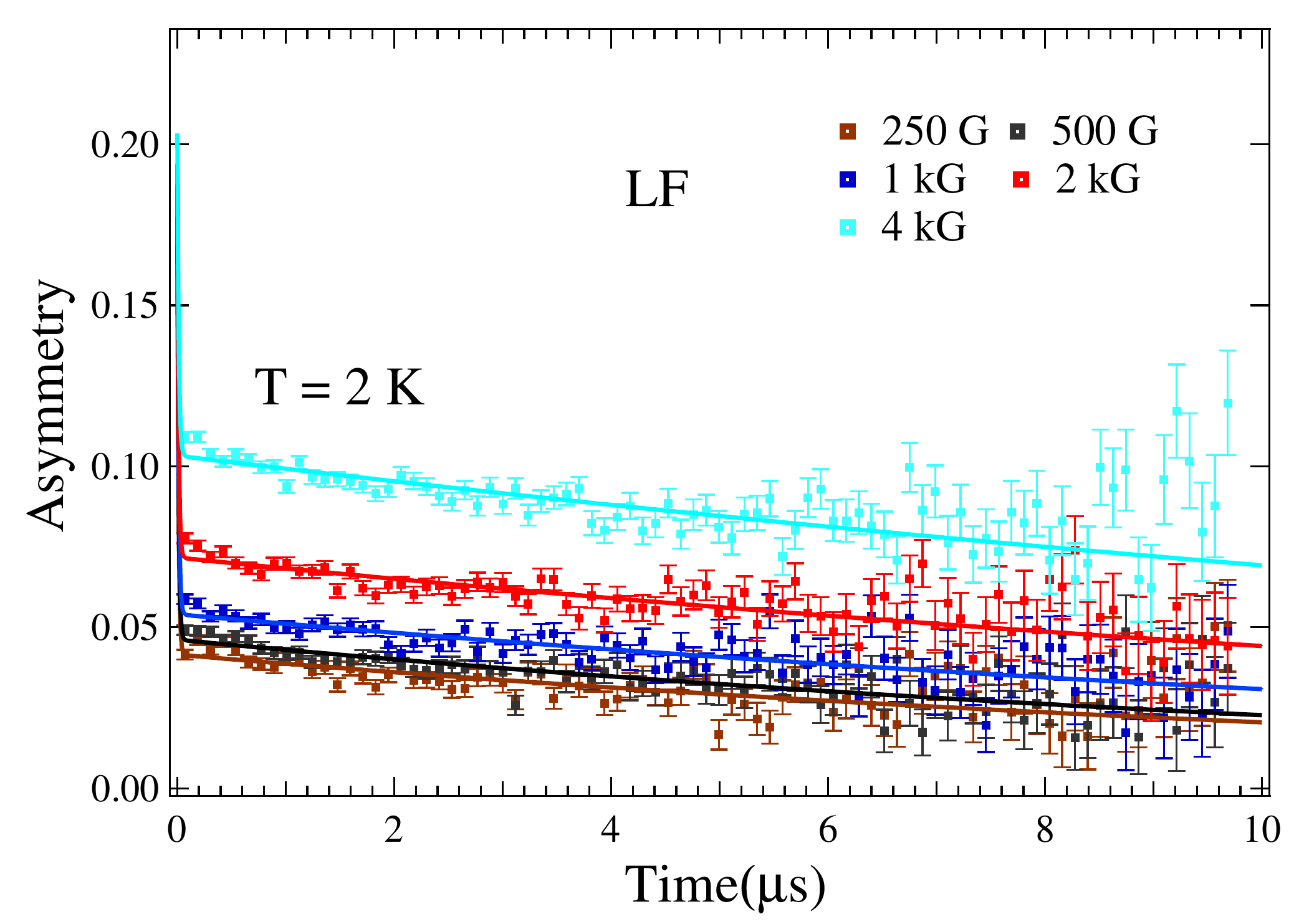}
\caption{\label{LFmusr}  Longitudinal field (LF) $\mu$SR data taken on polycrystalline sample of CuAlCr$_4$S$_8$ under different fields shows an upward shift of the 1/3rd tail.}
\end{figure}

We also performed longitudinal field (LF) $\mu$SR measurements. In longitudinal field $\mu$SR experiment, a field is applied parallel to the initial muon spin and this field cants the magnetic moments along the magnetic field, making the perpendicular field smaller, and effective parallel field larger, thereby shifting the 1/3rd tail upward for the polycrystalline samples. This process brings some of the fast relaxing asymmetry signal in the $\mu$SR time window. Figure \ref{LFmusr} shows the asymmetry spectra of the sample under different applied fields. The longitudinal field of 4kG shifts the $\mu$SR spectra up, however, it is not enough to bring the entire fast relaxing component of the signal in the $\mu$SR time window. This implies that the characteristic internal fields are quasi-static and are of the order of few thousand Gauss.

\section{Conclusion}
We successfully synthesized a new breathing pyrochlore sample CuAlCr$_4$S$_8$, and investigated the structural and magnetic properties. The specific heat data exhibits a sharp peak at 20 K, which is supported by an abrupt dip in the magnetic susceptibility. We observe a linear increase in magnetic transition at a rate of 0.53 K/kbar of applied helium pressure. The close proximity of breathing ratio, and $\theta_{CW}$ of CuAlCr$_4$S$_8$ to CuInCr$_4$S$_8$ may allow it to have similar complex nearest neighbor magnetic interactions, switching from antiferromagnetic in smaller tetrahedra to ferromagnetic in larger tertahedra; however, \textit{ab initio} calculations and neutron scattering experiments are required for more accurate determination of the exchange interaction parameters and magnetic ground state, respectively. 

\section{Acknowledgements}
We would like to thank Sarah Dunsiger for her assistance during the $\mu$SR measurements. Work at McMaster was supported by the Natural Sciences and Engineering Research Council of Canada.
Portions of this work are based upon research conducted at the Center for High Energy X-ray Sciences (CHEXS), which is supported by the National Science Foundation under award DMR-1829070. This research was undertaken thanks, in part, to funding from the Max Planck-UBC-UTokyo Center for Quantum Materials and the Canada First Research Excellence Fund, Quantum Materials and Future Technologies Program.


\nocite{*}

\bibliography{CuAlCr4S81}

\end{document}